\documentclass[conference,letterpaper]{IEEEtran}

\usepackage{cite}

\ifCLASSINFOpdf
\else
   \usepackage[dvips]{graphicx}
   \graphicspath{{../figures/}}
   \DeclareGraphicsExtensions{.eps}
\fi
\usepackage[cmex10]{amsmath}
\usepackage{amssymb,amsthm}
\usepackage{dsfont}

\theoremstyle{remark}

\newtheorem{lemma}{Lemma}
\newtheorem{theorem}{Theorem}
\newtheorem{corollary}{Corollary}
\newtheorem{definition}{Definition}
\newtheorem{identity}{Identity}

\usepackage{psfrag}
\usepackage{array}
\usepackage{mdwmath}
\usepackage{mdwtab}
\usepackage{enumerate}

\usepackage{eqparbox}
\usepackage{mathtools}

\usepackage{fixltx2e}

\usepackage{stfloats}

\begin{document}
%\title{Interference in Aloha-based Ad Hoc Networks with\\Isotropic Node Distribution and Rayleigh Fading}
\title{Interference and Throughput in Aloha-based\\Ad Hoc Networks with Isotropic Node Distribution}

\author{\IEEEauthorblockN{Ralph Tanbourgi, Holger J\"{a}kel, Leonid Chaichenets and Friedrich K. Jondral}
\IEEEauthorblockA{Communications Engineering Lab, Karlsruhe Institute of Technology, Germany\\
Email: \{ralph.tanbourgi, holger.jaekel, friedrich.jondral\}@kit.edu}
}

\maketitle

\begin{abstract}
%\boldmath
We study the interference and outage statistics in a slotted Aloha ad hoc network, where the spatial distribution of nodes is non-stationary and isotropic. In such a network, outage probability and local throughput depend on both the particular location in the network and the shape of the spatial distribution. We derive in closed-form certain distributional properties of the interference that are important for analyzing wireless networks as a function of the location and the spatial shape. Our results focus on path loss exponents $2$ and $4$, the former case not being analyzable before due to the stationarity assumption of the spatial node distribution. We propose two metrics for measuring local throughput in non-stationary networks and discuss how our findings can be applied to both analysis and optimization. 
\end{abstract}


% \fi
\IEEEpeerreviewmaketitle

%\footnote{The authors gratefully acknowledge that their work is partially
%supported within the priority program 1397 "COIN" under grant No. JO
%258/21-1 by the German Research Foundation (DFG).}
%\footnote{This work is partially supported within the priority program 1397 "COIN" under grant No. JO
%258/21-1 by the German Research Foundation (DFG).}
\section{Introduction}
The application of stochastic geometry to the modeling and analysis of wireless networks has attained a lot of attention in recent years. %Pioneered by the researchers of the DARPA IT-MANET program, stochastic geometry
It has enabled a new framework called \emph{transmission capacity} (TC) framework, which led to many new profound results in the topic of wireless networks (cf. \cite{weber10,ganti09}). The advantage of using a spatial model to describe the node positions rather than assuming a deterministic network topology is two-fold: First, such a probabilistic approach decouples the performance analysis from the \emph{actual} topology, thereby increasing the generality of results. Second, it provides powerful means for network optimization, especially for highly dynamic networks, where interference is (unpredictably) fast-varying. %In its definition, the TC is the maximum density of concurrent transmissions subject to an outage probability constraint, thereby implicitly performing a spatial averaging.

With few exceptions, the node positions are mostly modeled as a \emph{stationary} point process. Stationarity is a desirable property, allowing analytically tractable computations and, even more important, representing a key requirement for applying the definition of TC. Even though the stationarity assumption has not really narrowed the range of obtainable insights, it poses some shortcomings to the analysis of wireless networks:\\
\textbf{Infinite networks:} Stationarity implies that the network is infinitely large as opposed to real deployments with a finite number of nodes.\\
\textbf{No border effects:} Border effects are inherently neglected in infinite networks. However, border effects cause heterogeneity in the nodes' capabilities depending on their location, i.e., being dis-/connected, interference-/noise-limited, etc.\\
\textbf{Infinite interference for free-space path loss:} For stationary node distributions in the plane and a path loss exponent $\alpha=2$, the interference is infinite almost sure (a.s.) \cite{weber10}, resulting in a TC of zero. More specifically, stationary models lose their accuracy as the path loss exponent decreases due to the fact that infinitely many nodes contribute to the interference.\\
\textbf{Application of the TC:} As already mentioned, the TC applies only to stationary networks. When the node distribution is non-stationary, this metric must be modified to take into account heterogeneous node deployments.

In reality, wireless ad hoc networks always exhibit a heterogeneous node distribution. The most obvious example is perhaps when the nodes are distributed in a \emph{bounded} region. In such a network, the interference situation near the center will significantly differ from that at the border. Besides this simple example, more complex deployments are often found in practice, e.g., wireless sensor networks created by airdrop \cite{akyildiz02}, spontaneous formation of hot spots \cite{feeney01}, etc. 
\subsection{Contribution}
We extend the existing framework by relaxing the requirements on the node distribution, thereby allowing for \emph{isotropy only}. More specifically, we have the following results:
\begin{itemize}
	\item The interference and outage statistics for slotted Aloha with $\alpha=2$ and $\alpha=4$ are derived as a function of the receiver position and the spatial shape of the node distribution. % For suitably-chosen spatial shapes closed-form formulas are presented. 
	We consider a path loss plus block fading model. As for the outage statistics, we focus on Rayleigh fading. 
	We show how known results for the stationary case arise from our results as special cases.
	\item Two global metrics, namely the differential TC and the average sum throughput, that take into account heterogeneous node deployments are proposed. % and discussed. 
	While the former metric is a refinement of the TC, the latter quantifies the first order overall network efficiency. 
	%\item The analysis is supported by incorporating three channel models, namely path loss only, Rayleigh block fading and two-ray fading model, the latter recently attracting attention in the sensor networks community \cite{}. To further increase and demonstrate the practical relevance of this work, several optimization examples are presented.
\end{itemize}

\subsection{Related work}
Stationary models with heterogeneous node deployment %\footnote{In most cases, only the transmitter configuration is modeled, focusing on the resulting interference field.} 
have already been investigated. Specifically, Poisson-Cluster \cite{ganti09_1} and Mat\'{e}rn hard-core models \cite{baccelli09} have been studied, as they are well-suited for analyzing more sophisticated medium access control (MAC) schemes. Treated as \emph{general motion-invariant}, these and similar models were further analyzed in \cite{ganti11,ganti09,giacomelli11} in a unifying way. %Some finite networks may be constructed using cluster models and letting the \emph{parent} density tend to zero (with the representative cluster left). 
%Due to their stationarity However, the problem of infinite interference for $\alpha=2$ is not solved by this, as with all stationary models. 
In \cite{govindasamy11}, a non-stationary and isotropic node distribution was assumed for analyzing multi-antenna receivers. While the analysis showed that the shape of the spatial distribution has a considerable impact on link performance, the scenario was limited only to the case of the receiver located in the origin.
\section{Network model}\label{sec:model}
%We consider a wireless ad hoc network with nodes \emph{isotropically} distributed in the plane $\mathbb{R}^2$ according to a Poisson point process (PPP) $\Phi$ of intensity (density) $\lambda(x)$, where $x\in\mathbb{R}^2$. Due to the isotropy property of $\Phi$, $\lambda(x)$ is rotation-invariant and depends only on the Euclidean norm $\|x\|$, i.e., $\lambda(x)=\lambda(\|x\|,\phi)=\lambda(\|x\|)$, $\phi\in[0,2\pi)$. When working with polar coordinates, we will use the notation $\lambda(r)$, where $r:=\|x\|$.\footnote{We do not expect confusion when switching between different coordinate systems.} Using \cite{}, we can describe $\lambda(r)$ as the resulting intensity after applying a position-dependent thinning on a stationary PPP of intensity $\lambda$, i.e., 
% We consider a wireless ad hoc network with nodes \emph{isotropically} distributed in the plane $\mathbb{R}^2$ according to a point process (PP) $\Phi$ of intensity (density) $\lambda(x)$, where $x\in\mathbb{R}^2$. Due to the isotropy property of $\Phi$, $\lambda(x)$ is rotation-invariant and depends only on the Euclidean norm $\|x\|$, i.e., $\lambda(x)=\lambda(\|x\|,\phi)=\lambda(\|x\|)$, $\phi\in[0,2\pi)$. When working with polar coordinates, we will use the notation $\lambda(r)$, where $r:=\|x\|$.\footnote{We do not expect confusion when switching between different coordinate systems.} Using \cite{}, we can describe $\lambda(r)$ as the resulting intensity after applying a \emph{distance}-dependent thinning on a stationary PP of intensity $\lambda$, i.e., 
We consider a wireless ad hoc network with nodes \emph{isotropically} distributed in $\mathbb{R}^2$. The MAC employed by the nodes is slotted Aloha. In a randomly chosen slot, some nodes wish to transmit a packet. We assume that the set of transmitters $\{\mathsf{x}\}$ follows an isotropic Poisson point process (PPP) $\Phi_{\text{t}}:=\{\mathsf{x}\}$ on $\mathbb{R}^2$ with intensity $\lambda(x)$, where $x\in\mathbb{R}^2$. Due to the isotropy of $\Phi_{\text{t}}$, $\lambda(x)$ is rotation-invariant and depends only on the Euclidean norm $\|x\|$, i.e., $\lambda(x)=\lambda(\|x\|e^{j\phi})=\lambda(\|x\|)$, $\phi\in[0,2\pi)$. When working with polar coordinates, we will use the notation $\lambda(r)$, where $r:=\|x\|$. %\footnote{We do not expect confusion when switching between different coordinate systems.} 
With \cite{baccelli09}, we can describe $\lambda(r)$ as the resulting intensity after \emph{distance-dependent} thinning of a stationary PPP of intensity $\lambda$, i.e.,% $\lambda(r):=\lambda F(r)$,
\begin{IEEEeqnarray}{c}
	\lambda(r):=\lambda F(r),\IEEEeqnarraynumspace
\end{IEEEeqnarray}
where $F(r)$ is called the \emph{shape function} as it reflects the spatial shape of $\Phi_{\text{t}}$. We will pose the following restrictions on $F(r)$:
\begin{enumerate}[(i)]
	\item Positiveness: $F(r)\geq0$ for all $r\geq0$.
	\item Normalization: $\max_{r}\{F(r)\}=1$.
	%\item Rate of decay: $\lim_{r\to\infty} F(r)r^{\nu}<\infty$, for $\nu>2$.
\end{enumerate}
The restrictions (i) and (ii) are necessary to ensure that $\lambda(r)$ is non-negative and bounded by $\lambda$ everywhere. %(ii) and (iii) ensure that $\Phi_{\text{t}}$ has finite number of nodes a.s., since then $\int_{0}^{\infty}rF(r)\,\mathrm dr<\infty$. The intensity of $\Phi_{\text{t}}$ is now completely characterized by $\lambda$ and $F(r)$.

We assume that each transmitter $\mathsf{x}$ has an intended receiver $\mathsf{y}$ randomly located at fixed distance $d$. %FIXME Uniform angle distribution. 
From the random translation Theorem \cite{baccelli09} it follows that the set of receivers $\{\mathsf{y}\}$ forms an isotropic PPP $\Phi_{\text{r}}:=\{\mathsf{y}\}$ on $\mathbb{R}^2$ with intensity $\lambda(x)$ as well. 
The fixed distance assumption is commonly accepted, see \cite{weber10}. However, we will relax this assumption in Section \ref{sec:applications}.% to analyze border effects.%The combined set $\Phi_{\text{t}}\cup\Phi_{\text{r}}$ (transmitters and receivers) follows an isotropic non-Poisson PP.

%\begin{remark}
%	Note that $p_{\text{t}}+p_{\text{r}}\leq 1$ allows us to model also the fact that a fraction of $1-p_{\text{t}}-p_{\text{r}}$ nodes can turn into sleep-mode. From \cite{baccelli09} it follows that $\Phi_{\text{t}}$ and $\Phi_{\text{r}}$ are independent.
%\end{remark}
We consider a path loss plus block fading channel with independent and identically distributed (i.i.d.) fading coefficients. The power path loss between two positions $x,y\in\mathbb{R}^2$ is given by $\ell(\|x-y\|):=(c+\|x-y\|^{\alpha})^{-1}$ with path loss exponent $\alpha$. The parameter $c>0$ ensures boundedness of $\ell$. The power fading coefficient between a transmitter at $x$ and a receiver at $y$ is given by $\mathsf{g}_{xy}$, where $\mathbb{E}\left[\mathsf{g}_{xy}\right]=1$ for all $x,y\in\mathbb{R}^2$.

We further place a receiver at $y_{0}\in\mathbb{R}^2$ and an intended transmitter at an arbitrary position $x_{0}\in\mathbb{R}^2$ with distance $d$ to $y_{0}$. %Since $\Phi_{\text{r}}$ is isotropic, we can set $y_{0}\equiv (y_{0},0)$ in polar coordinates. 
The pair $x_{0}\to y_{0}$ is called the \emph{reference pair} as it will allow us to measure the (spatially-averaged) link performance for receivers at distance $\|y_{0}\|$ from the origin.% According to Slivnyak's Theorem \cite{}, the conditioning of $\Phi_{\text{t}}$ and $\Phi_{\text{r}}$ on having a transmitter at $x_{0}$ and a receiver at $y_{0}$, respectively, does not the change the distribution of $\Phi_{\text{t}}$ and $\Phi_{\text{r}}$. 

Assuming fixed power transmissions for all nodes, the instantaneous signal-to-interference-plus-noise ratio (SINR) at the reference receiver $y_{0}$ is given by:
\begin{IEEEeqnarray}{rCl}
	\mathsf{SINR}(y_{0})&:=&%\frac{\mathsf{g}_{x_{0}y_{0}}\ell(d)P}{N+\sum\limits_{ \mathclap{\mathsf{x}\in\Phi_{\text{t}}\setminus\{x_{0}\}}}\mathsf{g}_{\mathsf{x}y_{0}}\ell(\|\mathsf{x}-y_{0}\|)P}\IEEEeqnarraynumspace\IEEEnonumber\\
	\frac{\mathsf{g}_{x_{0}y_{0}}}{\eta+\ell(d)^{-1}\mathsf{I}(y_{0})},\IEEEeqnarraynumspace\label{eq:sinr}
\end{IEEEeqnarray}
where $\eta$ is the average noise-to-signal ratio and
\begin{IEEEeqnarray}{c}
	\mathsf{I}(y_{0}):=\sum\limits_{ \mathclap{\mathsf{x}\in\Phi_{\text{t}}\setminus\{x_{0}\}}}\mathsf{g}_{\mathsf{x}y_{0}}\ell(\|\mathsf{x}-y_{0}\|)
\end{IEEEeqnarray}
is the interference power. We assume strong channel coding, i.e., the outage event is a steep function of the SINR. The outage probability (OP) at the reference pair $x_{0}\to y_{0}$ is then given by the reduced Palm probability
\begin{IEEEeqnarray}{rCl}\label{eq:outage}
	q(y_{0}):=\mathbb{P}^{!x_{0}}\left(\mathsf{SINR}(y_{0})<\beta\right),
\end{IEEEeqnarray}
where $\beta$ is a modulation and coding specific SINR threshold.
%\begin{remark}
%	Since the nodes independently decide whether to transmit or not and because only the set of interferer contribute to the interference, it suffices to consider the probability measure with respect to $\Phi_{\text{t}}$, i.e., $\mathbb{P}_{\Phi_{\text{t}}}$.
%\end{remark}
\section{Interference analysis}\label{sec:interference}
\newcounter{tmpeqnnum}
\begin{figure*}[!b]
\normalsize
%\vspace*{4pt}
\hrulefill
\setcounter{tmpeqnnum}{\value{equation}}
\begin{IEEEeqnarray}{c}
\setcounter{equation}{12}
A_{4}(y_{0},c):=\frac{\pi}{2\sqrt{c}}\left(F(r)\,\text{arctan}\frac{2\text{Re}\{\kappa(r,c,y_{0})\}}{1-|\kappa(r,c,y_{0})|^2}
\bigg\vert_{r=0}^{\infty}-\int_{0}^{\infty} f(r)\,\text{arctan}\frac{2\text{Re}\{\kappa(r,c,y_{0})\}}{1-|\kappa(r,c,y_{0})|^2}\,\mathrm dr\right)\label{eq:A4}
\end{IEEEeqnarray}
\setcounter{equation}{\value{tmpeqnnum}}
\end{figure*}
We now study the interference statistics at the reference receiver at $y_{0}$. Before, we note the following two integral identities which are taken from \cite{gradshteyn07}:
\begin{identity}\label{lemma:integral_id1}
	 If $a>|b|$, $a,b\in\mathbb{R}$, 
	    \begin{IEEEeqnarray}{c}
			  \int_{0}^{\pi}\frac{\mathrm d\phi}{(a+b\cos\phi)^{n+1}}=\frac{\pi\,P_{n}\left(\frac{a}{\sqrt{a^2-b^2}}\right)}{(a^2-b^2)^{\frac{n+1}{2}}},\IEEEeqnarraynumspace\label{eq:integral_id1}
	    \end{IEEEeqnarray}
	    where $P_{n}(x)$ is the $n^{\text{th}}$-Legendre polynomial.
\end{identity}
%\begin{IEEEproof}
%	This correspondence can be found in \cite{gradshteyn07}.
%\end{IEEEproof}
\begin{identity}\label{lemma:integral_id2}
	Let $a_{1},a_{2},a_{3}\in\mathbb{R}$, $R:=a_{1}+a_{2}t^2+a_{3}t^4$, $\Delta=4a_{1}a_{3}-a_{2}^2$ and $a_{3}>0$. Using substitution $t\to t^2$, we have
	\begin{IEEEeqnarray}{c}
		\int\frac{2t\sqrt{a_{3}}\,\mathrm dt}{\sqrt{a_{1}+a_{2}t^2+a_{3}t^4}}=\begin{cases}
				\log\frac{2\sqrt{a_{3}R}+2a_{3}t^2+a_{2}}{\sqrt{\Delta}},& a_{3}>0\\
		            \text{arcsinh}\frac{2a_{3}t^2+a_{2}}{\sqrt{\Delta}}, & \Delta>0\\
		            \log(2a_{3}t^2+a_{2}), & \Delta=0.
		            \end{cases}\IEEEeqnarraynumspace\label{eq:integral_id2}
	\end{IEEEeqnarray}
\end{identity}
%\begin{IEEEproof}
%	The Lemma follows from \cite{gradshteyn07} after applying the substitution $t\to t^2$. 
%\end{IEEEproof}
We are now in the position to derive the first moment of the interference at $y_ {0}$.
\begin{theorem}\label{theorem:moment1}
Let $f(r):=\mathrm d F(r)/\mathrm d r$, $c>0$ and $\alpha=2$. If $\lim\limits_{r\to\infty}F(r)r^{\nu} < \infty$ for some $\nu>0$, then
	\begin{IEEEeqnarray}{c}
		\mathbb{E}^{!x_{0}}\left[\mathsf{I}(y_{0})\right] = \lambda A_{2}(y_{0},c)<\infty,\label{eq:exp_int2a}
	\end{IEEEeqnarray}
	where the \emph{interference-driving} function $A_{2}(y_{0},c)$ is given by
	\begin{IEEEeqnarray}{rCl}
		A_{2}(y_{0},c)&:=&F(0)\,\text{arcsinh}\frac{y_{0}^2-c}{2y_{0}\sqrt{c}}\IEEEnonumber\\
		&&+\int_{0}^{\infty}f(r)\,\text{arcsinh}\frac{y_{0}^2-r^2-c}{2y_{0}\sqrt{c}}\,\mathrm dr.\IEEEeqnarraynumspace\label{eq:A2}
	\end{IEEEeqnarray}
\end{theorem}
\begin{IEEEproof}
	We write
	\begin{IEEEeqnarray}{rCl}
		\mathbb{E}^{!x_{0}}\left[\mathsf{I}(y_{0})\right] &=& \lambda\int_{\mathbb{R}^2} \mathbb{E}\left[\mathsf{g}_{xy_{0}}\right]\ell(\|x-y_{0}\|) F(x) \,\mathrm dx\IEEEnonumber
		%&\overset{(b)}{=}&\lambda\pi\int_{0}^{\infty}\frac{2rF(r)\,\mathrm dr}{\sqrt{(c+r^2+y_{0}^2)^2-4r^2y_{0}^2}},\IEEEnonumber
	\end{IEEEeqnarray}
	what follows from Campbell's Theorem and Slivnyak's Theorem \cite{stoyan95}, and from the i.i.d. property of $\mathsf{g}_{\mathsf{x}y_{0}}$. Applying Identity \ref{lemma:integral_id1} and \ref{lemma:integral_id2} yields the result.
\end{IEEEproof}
The function $A_{2}(y_{0},c)$ in (\ref{eq:A2}) has an interesting interpretation: $A_{2}(y_{0},c)$ can be described as the interference field associated with the origin $o$, from which the remaining interference adds up differentially. %caused by transmitters in
\begin{corollary}
	Summary of some special cases of Theorem \ref{theorem:moment1}:
	\begin{enumerate}
		\item When we assume $F(0)=1$ and $f(r)\leq 0$ for all $r\in\mathbb{R}_{+}$, $F(r)$ can be interpreted as a complementary cumulative distribution function (CDF) with respect to a \emph{random} distance $\mathsf{r}$ to the origin, yielding
		\begin{IEEEeqnarray}{c}
			A_{2}(y_{0},c) = \text{arcsinh}\frac{y_{0}^2-c}{2y_{0}\sqrt{c}}-\mathbb{E}\left[\text{arcsinh}\frac{y_{0}^2-\mathsf{r}^2-c}{2y_{0}\sqrt{c}}\right].\IEEEnonumber
		\end{IEEEeqnarray}
		\item Letting $\|y_{0}\|\to 0$, we further have
		\begin{IEEEeqnarray}{c}
			A_{2}(0,c) = \log(1/2c)+\mathbb{E}\left[\log(2(\mathsf{r}+c))\right].\IEEEnonumber
		\end{IEEEeqnarray}
		\item Letting $c\to0$, we have $\mathbb{E}\left[\mathsf{I}(y_{0})\right]=\infty$, which is due to the resulting singularity of $\ell(\|x-y_{0}\|)$ at $x=y_{0}$, cf. \cite{ganti09}.
		\item Sparse network ($0<\lim_{r\to\infty}F(r)r^{\nu}<\infty$, $0<\nu\leq 2$): Remarkably, $\int_{0}^{\infty}rF(r)\mathrm dr=\infty$ but $\mathbb{E}^{!x_{0}}\left[\mathsf{I}(y_{0})\right]<\infty$.
		\item Dense network ($0<\lim_{r\to\infty}F(r)r^{\nu}<\infty$, $\nu\to0$): As expected \cite{ganti09}, $\mathbb{E}^{!x_{0}}\left[\mathsf{I}(y_{0})\right]=\infty$.
	\end{enumerate}
\end{corollary}
1) has an interesting interpretation as well: The expectation can be seen as \emph{averaging} the differential interference over $\mathsf{r}$. %Here, $\mathsf{r}$ is interpreted as the \emph{random} distance of an interferer to the origin, thereby implicitly reflecting the node distribution. 
Such an interpretation may be appropriate when analyzing networks with a priori unknown or fast-varying spatial configurations, for which a CDF is then used to model their spatial shape. 4) implies $\mathsf{I}(y_{0})<\infty$ a.s. although infinitely many nodes contribute to the interference on average. Note that 5) includes the homogeneous case with $F(r)=1$ ($f(r)=0$).

We now extend the findings of Theorem \ref{theorem:moment1}. Before, we need the following Lemma.
\begin{lemma}\label{lemma:pbz}
	Let $a_{1},a_{2}\in\mathbb{R}$, $a_{1}>0$. Then,
	\begin{IEEEeqnarray}{rCl}
	&&\hspace{-2cm}\int\int_{0}^{\pi}\frac{2t\,\mathrm d\phi\,\mathrm d t}{a_{1}+(t^2+a_{2}^2-2ta_{2}\cos\phi)^2}\IEEEnonumber\\
		%&=& \frac{\pi}{\sqrt{c}}\text{arctan}\frac{\sqrt{2}\left((r^2-y_{0}^2)\sqrt{|\kappa|+\text{Re}(\kappa)}-\sqrt{c}\sqrt{\|\kappa\|-\text{Re}(\kappa)}\right)}{\|-(r^2-y_{0}^2)^2-c}
		&\hspace{1cm}=&\frac{\pi}{2\sqrt{a_{1}}}\text{arctan}\frac{2\text{Re}\{\kappa(t,a_{1},a_{2})\}}{1-|\kappa(t,a_{1},a_{2})|^2},\IEEEeqnarraynumspace\label{eq:pbz}
	\end{IEEEeqnarray}
	where
	\begin{IEEEeqnarray}{c}
		\kappa(t,a_{1},a_{2}):=\frac{t^2-a_{2}^2-j\sqrt{a_{1}}}{\sqrt{(\sqrt{a_{1}}+j(t^2+a_{2}^2))^2+4t^2a_{2}^2}}.
	\end{IEEEeqnarray}
\end{lemma}
\begin{IEEEproof}
	The basic idea is to decompose the integrand into partial fractions and to apply Identity \ref{lemma:integral_id1} and \ref{lemma:integral_id2}, yielding (\ref{eq:pbz})  after some algebraic manipulations. Note that according to \cite{gradshteyn07}, (\ref{eq:integral_id1}) and (\ref{eq:integral_id2}) hold only for real-valued parameters. However, they were verified to hold also for complex-valued parameters. %Thus, we have
	%\begin{IEEEeqnarray}{c}
	%	\frac{1}{2\sqrt{c}}\sum\limits_{k=-1,1}\int t \int_{0}^{\pi}\frac{\mathrm d\phi \,\mathrm dt}{\sqrt{a_{1}}+kj(t^2+a_{2}^2-2ta_{2}\cos\phi)},\IEEEnonumber
	%\end{IEEEeqnarray}
	%where the inner integral can be calculated using Lemma \ref{lemma:integral_id1}. The result then follows after some algebraic manipulations and integration. 
\end{IEEEproof}
\begin{theorem}\label{theorem:moment2}
Let $f(r):=\mathrm d F(r)/\mathrm d r$, $c>0$ and $\alpha=4$. Then,
\begin{IEEEeqnarray}{c}
	\mathbb{E}^{!x_{0}}\left[\mathsf{I}(y_{0})\right] = \lambda\,A_{4}(y_{0},c)<\infty,\IEEEeqnarraynumspace\label{eq:exp_int2}
\end{IEEEeqnarray}
	where $A_{4}(y_{0},c)$ is given by (\ref{eq:A4}) below.
\end{theorem}
\setcounter{equation}{12}
\begin{IEEEproof}
	The proof is analogous to the proof of Theorem \ref{theorem:moment1} and uses the integral identity of Lemma \ref{lemma:pbz}. We further make use of (ii) in Section \ref{sec:model} to show that $\lim_{r\to\infty}F(r)<\infty$.
\end{IEEEproof}
\begin{corollary}\label{col:comp1}
	Summary of some special cases of Theorem \ref{theorem:moment2}:
	\begin{enumerate}
		\item Case $c\to0$: By taking the limit $\lim_{c\to0}A_{4}(y_{0},c)$ in (\ref{eq:A4}), we observe that $\mathbb{E}^{!x_{0}}\left[\mathsf{I}(y_{0})\right]=\infty$, cf. 3) in Corollary \ref{col:comp1}.
		%\item Infinite case ($0<\lim_{r\to\infty}F(r)<\infty$) and $c>0$: Here, we expect $\mathbb{E}^{!x_{0}}\left[\mathsf{I}(y_{0})\right]<\infty$. Indeed,
%			\begin{IEEEeqnarray}{c}
%				\lim\limits_{r\to\infty} F(r)\arctan\frac{2\text{Re}\{\kappa(r,c,y_{0})\}}{1-\|\kappa(r,c,y_{0})\|^2}=F(\infty)\,\frac{\pi}{2}<\infty.\IEEEnonumber
%			\end{IEEEeqnarray}
		\item Homogeneous case: Let $F(r)=1$. Then, $f(r)=0$ and % for all $r\in\mathbb{R}_{+}$ and the limits are
		\begin{IEEEeqnarray}{c}
			\lim\limits_{r\to a}\arctan\frac{2\text{Re}\{\kappa(r,c,y_{0})\}}{1-\kappa(r,c,y_{0})|^2}=
					\begin{cases}
						-\frac{\pi}{2}, & a=0\\
                   \frac{\pi}{2}, & a=\infty,
					\end{cases}\IEEEnonumber
		\end{IEEEeqnarray}
		yielding $\mathbb{E}^{!x_{0}}\left[\mathsf{I}(y_{0})\right]=\lambda\frac{\pi^2}{2\sqrt{c}}$ as expected, cf. \cite{ganti09}.
	\end{enumerate}
	All results of this Corollary are consistent with the literature.
\end{corollary}
The first moment of the interference is useful for bounding the interference distribution for the path loss only scenario. In case of Rayleigh fading channels, the Laplace transform of $\mathsf{I}(y_{0})$, i.e.,  $\mathcal{L}_{\mathsf{I}(y_{0})}(s):=\mathbb{E}\left[\exp\{-s\mathsf{I}(y_{0})\}\right]$, is of significant importance, since it allows one to obtain the OP in closed-form. When treating the case $\alpha=2$, we will always assume that $F(r)$ satisfies the additional condition of Theorem \ref{theorem:moment1}.  
\begin{theorem}\label{theorem:laplace_int}
	For $\mathsf{g}_{xy}\sim\text{Exp}(1)$ for all $x,y\in\mathbb{R}^2$ (Rayleigh fading), the Laplace transform of $\mathsf{I}(y_{0})$ at $y_{0}\in\mathbb{R}^2$ is
	\begin{IEEEeqnarray}{c}
		\mathcal{L}_{\mathsf{I}(y_{0})}(s)=\exp\left\{-\lambda s \,A_{\alpha}(y_{0},s+c)\right\},\IEEEeqnarraynumspace\label{eq:laplace}
	\end{IEEEeqnarray}
	for the cases $\alpha=2$ and $\alpha=4$, where $A_{2}(y_{0},c)$ is given by (\ref{eq:A2}) and $A_{4}(y_{0},c)$ is given by (\ref{eq:A4}).
\end{theorem}
\begin{IEEEproof}
	We write
	\begin{IEEEeqnarray}{rCl}
	\mathcal{L}_{\mathsf{I}(y_{0})}(s)&\overset{(a)}{=}&\mathbb{E}^{!x_{0}}_{\Phi_{\text{t}}}\left[\prod\limits_{\mathsf{x}\in\Phi_{\text{t}}} \mathbb{E}_{\mathsf{g}_{\mathsf{x}y_{0}}}\left[\exp\left\{-s\mathsf{g}_{\mathsf{x}y_{0}}\ell(\|\mathsf{x}-y_{0}\|)\right\}\right]\right]\IEEEnonumber\\
		&\overset{(b)}{=}&\exp\left\{-\int_{\mathbb{R}^2}  \left(1-\mathcal{L}_{\mathsf{g}}\left(s\ell(\|x-y_{0}\|)\right)\right)\,\lambda(x)\mathrm dx\right\},\IEEEnonumber
	\end{IEEEeqnarray}
	where (a) follows from algebraic manipulations and the i.i.d. property of the $\mathsf{g}_{\mathsf{x}y_{0}}$. (b) follows from the probability generating functional and the Laplace functional of a PPP \cite{baccelli09}. After noting that $\mathcal{L}_{\mathsf{g}}(s)=(1+s)^{-1}$ for $\mathsf{g}\sim\text{Exp}(1)$, the integral is computed using Identity \ref{lemma:integral_id1} and \ref{lemma:integral_id2}, and Lemma \ref{lemma:pbz}.
\end{IEEEproof}
Note that (a) in the proof holds for general point processes and some approximation techniques for computing the right-hand side already exist \cite{ganti10}. The (b) part is for PPPs only.

\begin{corollary}\label{col:comp2}
Setting $F(r)=1$ for all $r\in\mathbb{R}_{+}$ and $c=0$, we obtain the well-known result for the homogeneous case with $\alpha=4$ \cite{ganti09}: $\mathcal{L}_{\mathsf{I}(y_{0})}(s)=\exp\{-\lambda\frac{\pi^2}{2} \sqrt{s}\}$.
\end{corollary}

\section{Outage and Local Throughput}
\subsection{Outage probability}
\begin{figure}[t]
	\psfrag{tag1}[c][c]{\small{$\eta=0$}}
	\psfrag{tag2}[c][c]{\small{$\eta=0.1$}}
	\psfrag{tag3}[c][c]{\small{$\|y_{0}\|$}}
	\psfrag{tag4}[c][c]{\small{$q(y_{0})$}}
	\psfrag{tag5tag5}{\footnotesize{$\alpha=2$}}
	\psfrag{tag6}{\footnotesize{$\alpha=4$}}
		   \centering
    \includegraphics[width=0.47\textwidth]{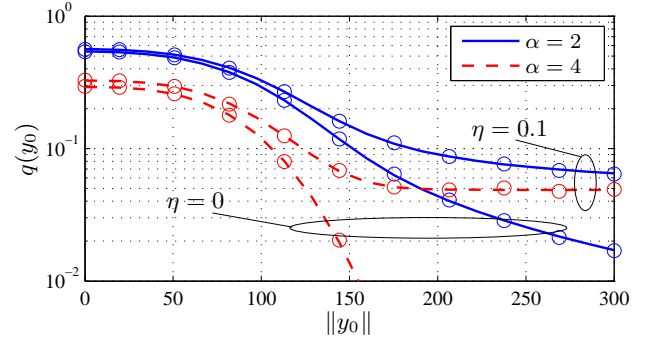}
\caption{$q(y_{0})$ vs. $\|y_{0}\|$ for $F(r)=\exp\{-(r/100)^{3}\}$, $d=10$, $\beta=0.5$, $c=1$, $\lambda=0.001$. Marks represent the simulation results.}
\label{fig:op}\vspace{-0.2cm}
\end{figure}

We now study the OP for the reference pair $x_{0}\to y_{0}$. In order to broadly discuss the impact of the spatial shape on the performance, we focus on the Rayleigh fading scenario. For other channel models, the interference moments derived in Section \ref{sec:interference} can be used to effectively bound the OP, e.g., using the Markov inequality \cite{weber10}. We do not expect additional insights by considering also other channel models.
\begin{theorem}\label{theorem:op}
The OP for the Rayleigh fading scenario and $\alpha=2$ respectively $\alpha=4$ is given by
	\begin{IEEEeqnarray}{c}
		q(y_{0})=1-\mathcal{L}_{\mathsf{I}(y_{0})}\left(\beta(c+d^{\alpha})\right)\exp\left\{-\beta\eta\right\}.\IEEEeqnarraynumspace\label{eq:op}
	\end{IEEEeqnarray}
\end{theorem}
\begin{IEEEproof}
	It is well-known that the OP for Aloha MAC and exponentially distributed power gains $\mathsf{g}_{xy}$ can be written in terms of the Laplace transform of the interference \cite{baccelli09,ganti09}: We condition (\ref{eq:sinr}) on $\Phi$ and evaluate the OP first with respect to $\mathsf{g}_{x_{0}y_{0}}$. We finally use (\ref{eq:laplace}) with $s=\beta(c+d^{\alpha})$.
\end{IEEEproof}
By means of (\ref{eq:op}) in Theorem \ref{theorem:op} we can now measure the OP for Rayleigh fading at an arbitrary location for an arbitrary spatial shape function $F(r)$ satisfying the given restrictions. Fig. \ref{fig:op} shows $q(\|y_{0}\|)$ vs. $\|y_{0}\|$ for $\alpha=2$ and $\alpha=4$, thereby confirming the analysis. It can further be observed how the network ``moves'' from the interference-limited to the power-limited regime with increasing $\|y_{0}\|$.

\begin{figure}[t]
	\psfrag{tagx}[c][c]{\small{$\|y_{0}\|$}}
	\psfrag{tagy1}[c][c]{\small{$\gamma$}}
	\psfrag{tagy2}[c][c]{\small{$F(\|y_{0}\|)$}}
	\psfrag{tag3tag3}{\footnotesize{$\gamma$}}
	\psfrag{data2}{\footnotesize{$F(\|y_{0}\|)$}}
		   \centering
    \includegraphics[width=0.47\textwidth]{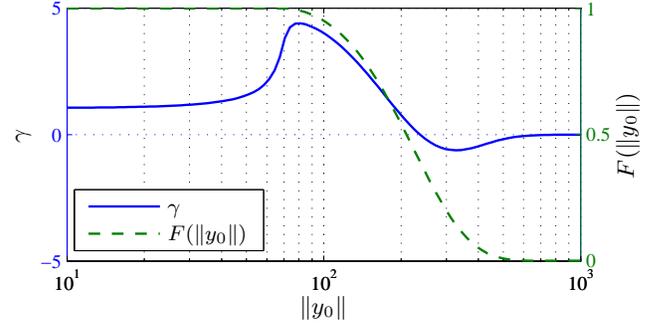}
\caption{$\gamma$ and $F(\|y_{0}\|)$ vs. $\|y_{0}\|$ for $\alpha=4$, $\beta=1$, $d=10$, $c=1$.}\vspace{-0.45cm}
\label{fig:gamma}
\end{figure}

To highlight the accuracy of the model, we compare the OP from Theorem \ref{theorem:op} to a straightforward way of approximating the OP which consists of assuming that the intensity $\lambda(x)$ is approximately constant around $y_{0}$. In this case the OP can then be described as in the homogeneous case \cite{ganti09}, except for the intensity in the exponential term being modulated by $F(y_{0})$, i.e., $\tilde{q}(y_{0}):=1-\exp\{-F(y_{0})\lambda\pi^2 d^2\beta^{\frac{2}{\alpha}}\frac{2}{\alpha}\csc\frac{2\pi}{\alpha}\}\approx q(y_{0})$. We will now study the logarithmic ratio of exact to approximate success probability, i.e., $\gamma:=\log\frac{1-q(y_{0})}{1-\tilde{q}(y_{0})}$.
\begin{corollary}
	Let $c=0$. The ratio $\gamma$ for $\alpha=4$ is given by
	\begin{IEEEeqnarray}{c}
		\gamma=\lambda d^{2}\sqrt{\beta}\left(\tfrac{\pi^2}{2}F(y_{0})-d^{2}\sqrt{\beta} A_{4}(y_{0},\beta d^{4})\right).\IEEEeqnarraynumspace
	\end{IEEEeqnarray}
\end{corollary}
Fig. \ref{fig:gamma} shows the ratio $\gamma$ together with the shape function $F(r)$ for different receiver positions $y_{0}$. $F(r)$ was chosen such that the network exhibits a communication hotspot, with the density of active nodes slowly decaying between $70$ and $500$ until it becomes approximately zero. One can see that the approximation is not satisfactory, especially in the transition region, where border effects come into play.
\subsection{Local throughput}
We now propose two local throughout metrics that are suitable for non-stationary wireless ad hoc networks.
\begin{definition}[Differential transmission capacity (DTC)]
	The DTC is defined as the maximal density of concurrent transmissions in an \emph{infinitesimal} region around the point $x\in\mathbb{R}^2$ subject to an OP constraint $\epsilon$, i.e.,
	\begin{IEEEeqnarray}{c}
		c(x,\epsilon):=\lambda(x,\epsilon)(1-\epsilon).\label{eq:def_dtc}
	\end{IEEEeqnarray}
\end{definition}
The TC and its differential counterpart have similar meaning, except that the latter is \emph{position-dependent}: For a given spatial shape $F(r)$ and target OP $\epsilon$, $c(x,\epsilon)$ yields the TC in a region $\mathrm dx$. Hence, the DTC implicitly takes into account the spatial shape of the node distribution. For Rayleigh fading, $c(x,\epsilon)$ is obtained by solving (\ref{eq:op}) for $\lambda$. Like the TC, the DTC can be used for comparing different transmission protocols.
\begin{definition}[Average sum throughput (AST)]
		The AST is defined as the ratio of average number of successful transmissions to average number of simultaneous transmissions, i.e.,
		\begin{IEEEeqnarray}{c}
		\Omega:=\frac{\mathbb{E}\Big[\sum\limits_{\mathsf{x}\in\Phi_{\text{t}}}\mathds{1}_{\{\mathsf{x}\text{ successful}\}}\Big]}{\mathbb{E}\Big[\sum\limits_{\mathsf{x}\in\Phi_{\text{t}}}\mathds{1}_{\{\mathsf{x}\in\mathbb{R}^2\}}\Big]}.
		\end{IEEEeqnarray}
		%where the averaging is over the transmitter set $\Phi_{\text{t}}$.
\end{definition}

The AST quantifies the first order overall efficiency of the network on the MAC layer. While the DTC highlights the spatial dynamics of the local throughput, the AST yields a single figure of merit. In essence, the AST counts the number of successful transmissions, thereby integrating over the spatial dynamics. Note that the success function $\mathds{1}_{\{\mathsf{x}\text{ successful}\}}$, indicating that transmitter $\mathsf{x}$ has been successful, can be chosen arbitrarily to include additional outage-inducing effects, e.g., energy-limitations, dis-connectivity or secrecy outage.
\begin{theorem}\label{theorem:fc}
	Let $\lim_{r\to\infty}F(r)r^{\nu}<\infty$ for some $\nu>2$. With the underlying network model and success function $\mathds{1}_{\{\mathsf{SINR}(y)\geq\beta\}}$, the AST $\Omega$ can be computed as\vspace{-0.1cm}
	\begin{IEEEeqnarray}{c}
		\Omega=\frac{\int_{0}^{\infty}r(1-q(r))\,F(r)\,\mathrm dr}{\int_{0}^{\infty}r F(r)\,\mathrm dr}.
	\end{IEEEeqnarray}
\end{theorem}
\begin{IEEEproof}
	Since the denominator directly follows from Campbell's Theorem, we focus on the numerator and write
	\begin{IEEEeqnarray}{rCl}
	&&\mathbb{E}\left[\sum\limits_{\mathsf{x}\in\Phi_{\text{t}}}\mathds{1}_{\{\mathsf{x}\text{ successful}\}}\right]\overset{(a)}{=}\int_{\mathbb{R}^2} \mathbb{E}^{!x} \left[\mathds{1}_{\{x\text{ successful}\}}\right]\lambda(x)\,\mathrm dx\IEEEnonumber\\
	&&\overset{(b)}{=}\int_{\mathbb{R}^2}\int_{\mathbb{R}^2} \mathbb{E}^{!x}_{\Phi_{\text{t}}} \left[\mathds{1}_{\{\mathsf{SINR}(y)\geq\beta\}}\right]\mathbb{P}\left(\mathsf{y}=y|x\right)\,\mathrm dy\,\lambda(x)\,\mathrm dx\IEEEnonumber\\
	&&\overset{(c)}{=}\int_{\mathbb{R}^2}\left(\int_{\mathbb{R}^2} \mathbb{E}^{!x}_{\Phi_{\text{t}}} \left[\mathds{1}_{\{\mathsf{SINR}(y)\geq\beta\}}\right]\mathbb{P}\left(\mathsf{y}=y|x\right)\lambda(x)\,\mathrm dx\right)\mathrm dy\IEEEnonumber\\
	&&\overset{(d)}{=}\int_{\mathbb{R}^2}\left(\int_{\mathbb{R}^2} \mathbb{P}^{!x}_{\Phi_{\text{t}}} \left(\mathsf{SINR}(y)\geq\beta\right)\mathbb{P}\left(\mathsf{y}=y|x\right)\,\lambda(x)\,\mathrm dx\right)\mathrm dy\IEEEnonumber\\
	&&\overset{(e)}{=}\int_{\mathbb{R}^2}(1-q(y))\left(\int_{\mathbb{R}^2} \mathbb{P}\left(\mathsf{y}=y|x\right)\,\lambda(x)\,\mathrm dx\right)\mathrm dy\IEEEnonumber\\
	&&=\int_{\mathbb{R}^2}(1-q(y))\lambda(y)\,\mathrm dy.\IEEEnonumber
	% &\overset{(b)}{=}& \frac{p\int_{\mathbb{R}^2} \mathbb{P}^{!x_{0}}_{\Phi_{\text{t}}} \left(\mathsf{SINR}(y_{0})\geq\beta\right)\lambda(x)\,\mathrm dx}{2p\lambda\pi\int_{0}^{\infty}r F(r)\,\mathrm dr},
	\end{IEEEeqnarray}
(a) is due to Campbell's Theorem \cite{stoyan95}. (b) is obtained by noting that a transmitter $x$ is successful if the intended receiver at $y$ is not in outage. From Section \ref{sec:model}, we know that $y$ is placed by random translation of $x$ according to some probability kernel $\mathbb{P}\left(\mathsf{y}=y|x\right)$. (c) follows from Tonelli's Theorem \cite{bauer92} and (d) follows from $\mathbb{E}\left[\mathds{1}_{\{X\in A\}}\right]=\mathbb{P}\left(X\in A\right)$. (e) follows from (\ref{eq:op}) and the fact that $q(y)$ is independent of $x$.
\end{IEEEproof}
\section{Applications of the Model}\label{sec:applications}%\vspace{-.1cm}
%We now discuss the practical use of our model.
\subsection{Shot-range inhibition}
Besides slotted Aloha, other MAC protocols such as CSMA/CA or local FDMA, are promising techniques for reducing excessive interference generated by nodes within \emph{shot-range}. To study ad hoc networks with such inhibition mechanisms while ensuring analytical tractability, powerful methods based on non-homogeneous Poisson approximation have been used \cite{hunter10,baccelli09,tanbourgi11_2}. When such protocols are \emph{transmitter-initiated}, e.g., transmitter sensing for CSMA or transmitter orthogonalization for FDMA, the resulting spatial distribution of interferers becomes inhomogeneous and approximately isotropic around the transmitter $x$, while the interference field at the intended receiver $y$ will depend on the distance $\|x-y\|$. Hence, our model can be applied also to such modeling problems and is not limited to Aloha MAC. 
\subsection{Network optimization}
\begin{figure}[t]
	\psfrag{tag4}[c][c]{\small{$\beta$ dB}}
	\psfrag{tag5}[c][c]{\small{$\log_{2}(1+\beta)\,\Omega(\beta)$}}
	\psfrag{tag1tag1tag1tag1tag1tag1tag1tag1tag}{\tiny{$\eta_{10}=-8$ dB, $\lambda=10^{-3}$, $\lambda_{\text{r}}=10^{-2}$}}
	\psfrag{tag2}{\tiny{$\eta_{10}=-14$ dB, $\lambda=10^{-2}$, $\lambda_{\text{r}}=10^{-2}$}}
	\psfrag{tag3}{\tiny{$\eta_{10}=-18$ dB, $\lambda=3\cdot 10^{-4}$, $\lambda_{\text{r}}=10^{-3}$}}
		   \centering
    \includegraphics[width=0.47\textwidth]{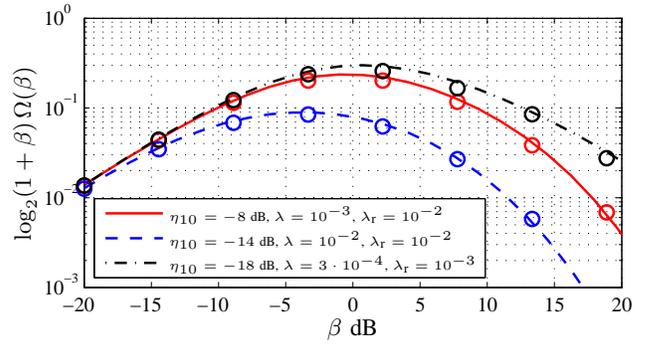}
\caption{$\log_{2}(1+\beta)\,\Omega(\beta)$ vs. $\beta$ for $\alpha=2$. $\eta_{10}$ denotes $\eta$ at a distance $d=10$. $F(r)=\exp\{-r/250\}$. Marks represent simulation results.}
\label{fig:opt}\vspace{-0.4cm}
\end{figure}

%In order to demonstrate how the analysis can be used for optimization problems,
Consider the following situation: Let the set $\{\mathsf{y}\}$ of \emph{potential} receivers be distributed as an isotropic PPP $\Phi_{\text{r}}$ of intensity $\lambda_{\text{r}}(r)=\lambda_{\text{r}} F(r)$. Assume that $\Phi_{\text{t}}$ and $\Phi_{\text{r}}$ are independent. That is, the set of all nodes follows a PPP, e.g., a sensor network created by airdrop, and connectivity at distance $d$ is no longer guaranteed for every node. We further assume that the routing protocol employs a nearest neighbor strategy, i.e., transmitters aim at minimizing $d$. For points distributed as a PPP, the CDF $F_{\mathsf{d}}(d)$ of the distance $\mathsf{d}$ between a point and its nearest neighbor is well-known, see \cite{baccelli09}. We would like to know the optimal SINR threshold $\beta$ such that the product $\log_{2}(1+\beta)\,\Omega(\beta)$ with success function $\mathds{1}_{\{\mathsf{x}\text{ successful}\}}\mathds{1}_{\{\mathsf{x}\text{ connected}\}}$ is maximized. This corresponds to maximizing the expected \emph{sum rate}, i.e.,
\begin{IEEEeqnarray}{rCl}
		\beta^{\ast}&=&\arg\max\limits_{\beta}\left\{\log_{2}(1+\beta)\,\Omega(\beta)\right\}\IEEEnonumber\\
		&\overset{(a)}{\approx}& \arg\max\limits_{\beta}\left\{\log_{2}(1+\beta)\hspace{-0.1cm}\int_{0}^{\infty}\hspace{-0.3cm} r \hspace{-0.1cm}\int_{0}^{\infty}\hspace{-0.3cm} (1-q(r,\beta,d))F_{\mathsf{d}}(\mathrm dd) \mathrm dr\right\}\hspace{-0.1cm},\label{eq:opt}\IEEEnonumber
\end{IEEEeqnarray}
where we altered the notation $\Omega\,{\to}\,\Omega(\beta)$ and $q(r)\,{\to}\, q(r,\beta,d)$ to point out the functional dependencies. (a) follows from
\begin{IEEEeqnarray}{c}
\mathbb{E}\left[q(\|\mathsf{y}_{i}\|)|\mathsf{x}_{i}=x_{i}\right]%\approx q\left(\mathbb{E}\left[\|\mathsf{y}_{i}\|\,|\mathsf{x}_{i}=x_{i}\right]\right)
\approx q(\|x_{i}\|),\IEEEnonumber
\end{IEEEeqnarray}
which essentially approximates the interference field at a receiver $y_{i}$ by the interference field at the associated transmitter $x_{i}$. This approximation is reasonable for high $\lambda_{\text{r}}$ and/or moderate slopes of $F(r)$. Fig. \ref{fig:opt} shows $\log_{2}(1+\beta)\,\Omega(\beta)$ vs. $\beta$. As can be seen, optimizing over $\beta$ yields large improvements. %The optimization problem can be reformulated as a spatial-aware optimization problem, i.e., after node deployment, by optimization over the functions $\beta(r)$.

\section{Concluding Remarks}
We extended prior work on the modeling and analysis of wireless networks by assuming an isotropic but not necessarily stationary spatial distribution of nodes. We derived, for slotted Aloha, the interference and outage statistics as a function of the receiver position and the shape of the spatial node distribution. The case $\alpha=2$, which could not be studied yet due to the stationarity assumption, was intensively studied. For $\alpha=4$, we also obtain closed-form results, from which known results arise as special cases. We proposed two metrics for measuring local throughput in non-stationary and finite networks and discussed possible applications of our model.

\vspace{-0.00cm}
\section*{Acknowledgements}
The authors gratefully acknowledge that their work is partially
supported within the priority program 1397 "COIN" under grant No. JO
258/21-1 by the German Research Foundation (DFG).
\vspace{-0.3cm}





%\bibliographystyle{IEEEtran}
\bibliographystyle{IEEEtran}
%\bibliography{IEEEabrv,../../literature}

\begin{thebibliography}{10}
\providecommand{\url}[1]{#1}
\csname url@samestyle\endcsname
\providecommand{\newblock}{\relax}
\providecommand{\bibinfo}[2]{#2}
\providecommand{\BIBentrySTDinterwordspacing}{\spaceskip=0pt\relax}
\providecommand{\BIBentryALTinterwordstretchfactor}{4}
\providecommand{\BIBentryALTinterwordspacing}{\spaceskip=\fontdimen2\font plus
\BIBentryALTinterwordstretchfactor\fontdimen3\font minus
  \fontdimen4\font\relax}
\providecommand{\BIBforeignlanguage}[2]{{%
\expandafter\ifx\csname l@#1\endcsname\relax
\typeout{** WARNING: IEEEtran.bst: No hyphenation pattern has been}%
\typeout{** loaded for the language `#1'. Using the pattern for}%
\typeout{** the default language instead.}%
\else
\language=\csname l@#1\endcsname
\fi
#2}}
\providecommand{\BIBdecl}{\relax}
\BIBdecl

\bibitem{weber10}
S.~Weber, J.~Andrews, and N.~Jindal, ``An overview of the transmission capacity
  of wireless networks,'' \emph{IEEE Trans. Commun.}, Dec. 2010.

\bibitem{ganti09}
M.~Haenggi and R.~K. Ganti, ``Interference in large wireless networks,''
  \emph{Found. Trends Netw.}, vol.~3, pp. 127--248, Feb. 2009.

\bibitem{akyildiz02}
I.~Akyildiz, W.~Su, Y.~Sankarasubramaniam, and E.~Cayirci, ``A survey on sensor
  networks,'' \emph{IEEE Commun. Magazine}, pp. 102--114, Aug. 2002.

\bibitem{feeney01}
L.~Feeney, B.~Ahlgren, and A.~Westerlund, ``Spontaneous networking: an
  application oriented approach to ad hoc networking,'' \emph{IEEE Commun.
  Magazine}, vol.~39, no.~6, pp. 176 --181, June 2001.

\bibitem{ganti09_1}
R.~Ganti and M.~Haenggi, ``Interference and outage in clustered wireless ad hoc
  networks,'' \emph{IEEE Trans. Inf. Theory}, vol.~55, no.~9, Sep. 2009.

\bibitem{baccelli09}
F.~Baccelli and B.~Blaszczyszyn, ``Stochastic geometry and wireless networks,
  volume 1+2: Theory and applications,'' \emph{Foundations and Trends in
  Networking}, 2009.

\bibitem{ganti11}
R.~Ganti, J.~Andrews, and M.~Haenggi, ``High-sir transmission capacity of
  wireless networks with general fading and node distribution,'' \emph{IEEE
  Trans. Inf. Theory}, vol.~57, no.~5, pp. 3100 --3116, May 2011.

\bibitem{giacomelli11}
R.~Giacomelli, R.~Ganti, and M.~Haenggi, ``Outage probability of general ad hoc
  networks in the high-reliability regime,'' \emph{IEEE/ACM Trans. Networking},
  vol.~19, no.~4, pp. 1151 --1163, Aug. 2011.

\bibitem{govindasamy11}
S.~Govindasamy and D.~Bliss, ``On the spectral efficiency of links with
  multi-antenna receivers in non-homogenous wireless networks,'' in \emph{IEEE
  Int. Conf. Commun. (ICC)}, June 2011.

\bibitem{gradshteyn07}
I.~S. Gradshteyn and I.~M. Ryzhik, \emph{Table of integrals, series, and
  products}, 7th~ed.\hskip 1em plus 0.5em minus 0.4em\relax Elsevier/Academic
  Press, Amsterdam, 2007.

\bibitem{stoyan95}
D.~Stoyan, W.~Kendall, and J.~Mecke, \emph{Stochastic geometry and its
  applications}, 2nd~ed.\hskip 1em plus 0.5em minus 0.4em\relax Wiley, 1995.

\bibitem{ganti10}
R.~Ganti and J.~Andrews, ``A new method for computing the transmission capacity
  of non-poisson wireless networks,'' in \emph{Proc. IEEE Int. Symposium on
  Information Theory (ISIT)}, July 2010.

\bibitem{bauer92}
H.~Bauer, \emph{Mass- und Integrationstheorie}, 2nd~ed., ser. De Gruyter
  Lehrbuch.\hskip 1em plus 0.5em minus 0.4em\relax Berlin: de Gruyter, 1992.

\bibitem{hunter10}
A.~Hunter, R.~Ganti, and J.~Andrews, ``Transmission capacity of multi-antenna
  ad hoc networks with csma,'' in \emph{Forty Fourth Asilomar Conf. on Signals,
  Systems and Computers (ASILOMAR)}, Nov. 2010.

\bibitem{tanbourgi11_2}
R.~Tanbourgi, J.~Elsner, H.~J\"{a}kel, and F.~Jondral, ``Lower bounds on the
  success probability for ad hoc networks with local fdma scheduling,'' in
  \emph{Workshop on Spatial Stochastic Models for Wireless Networks (SpaSWiN)},
  May 2011.

\end{thebibliography}

\end{document}